\documentclass[10pt,conference,letterpaper]{RWWTemplate}

\usepackage{times,amsmath,epsfig}
\usepackage{cite}
\usepackage{amsmath,amssymb,amsfonts}
\usepackage{capt-of}
\usepackage{graphicx}
\usepackage{textcomp}
\usepackage{xcolor}
\usepackage[draft]{hyperref}
\usepackage{multirow}
\usepackage{float}
\usepackage{wrapfig}
\usepackage{hyperref}
\usepackage{caption}
\usepackage{subcaption}
\usepackage{epstopdf}
\usepackage{color}
\usepackage{longtable}
\usepackage{moreverb}
\usepackage{algorithmicx}
\usepackage{algorithm}
\usepackage{algpseudocode}
\usepackage[labelsep=period]{caption}
\usepackage{textcase}
\usepackage{mathtools}
\usepackage{enumerate}

\renewcommand{\baselinestretch}{0.98}
\title{Interference Coordination for Aerial and Terrestrial Nodes in Three-Tier LTE-Advanced HetNet}

\author{%
Abhaykumar Kumbhar$^{1}$, Hamidullah Binol$^{1}$, \.{I}smail~G\"uven\c{c}$^{2}$, and Kemal Akkaya$^{1}$ %
\vspace{12pt}\\
$^1$Dept. Electrical and Computer Engineering, Florida International University, Miami, FL, 33174\\
$^2$Dept. Electrical and Computer Engineering, North Carolina State University, Raleigh, NC, 27606 \\
}

\begin{document}
\maketitle

%
\begin{abstract}
Integrating unmanned aerial vehicles (UAVs) as user equipment (UE) and base-stations (BSs) into an existing LTE-Advanced heterogeneous network (HetNet) can further enhance wireless connectivity and support emerging services. However, this would require effective configuration of system-level design parameters for interference management. This paper provides system-level insights into a three-tier LTE-Advanced air/ground HetNet, wherein the UAVs are deployed both as BSs and UEs, and co-exist with existing terrestrial nodes. Moreover, this HetNet leverages on cell range expansion (CRE), intercell interference coordination (ICIC), 3D beamforming, and enhanced support for UAVs. Through Monte-Carlo simulations, we compare system-wide fifth percentile spectral efficiency (5pSE) and coverage probability for different ICIC techniques, while jointly optimizing the ICIC and CRE parameters. Our results show that reduced power subframes defined in 3GPP Rel-11 can provide considerably better 5pSE and coverage probability than the 3GPP Rel-10 with almost blank subframes.
\end{abstract}


\let\thefootnote\relax\footnotetext{This research was supported in part by NSF under CNS-1453678.}
\section{Introduction}
Several of the telecommunications service providers are considering the use of unmanned aerial vehicles (UAVs) to meet the mobile data and coverage demands, restore damaged infrastructure, and enable emerging service~\cite{R1,chandrasekharan2016designing}. However, integration of these UAVs as aerial user equipment (AUEs) and unmanned aerial base-stations (UABSs), would require a system-level understanding to both modify and extend the existing terrestrial network infrastructure. A vital goal while planning any air/ground heterogeneous network (AG-HetNet) is to ensure ubiquitous data coverage with broadband rates. To this end, existing works have explored the co-existence of terrestrial and aerial nodes in a network and assessed the performance this AG-HetNet in terms of \textit{coverage probability} and \textit{fifth percentile spectral efficiency} (5pSE) as the two key performance indicators (KPIs).

\begin{figure} [t]
\centering
\includegraphics[width=0.95\linewidth]{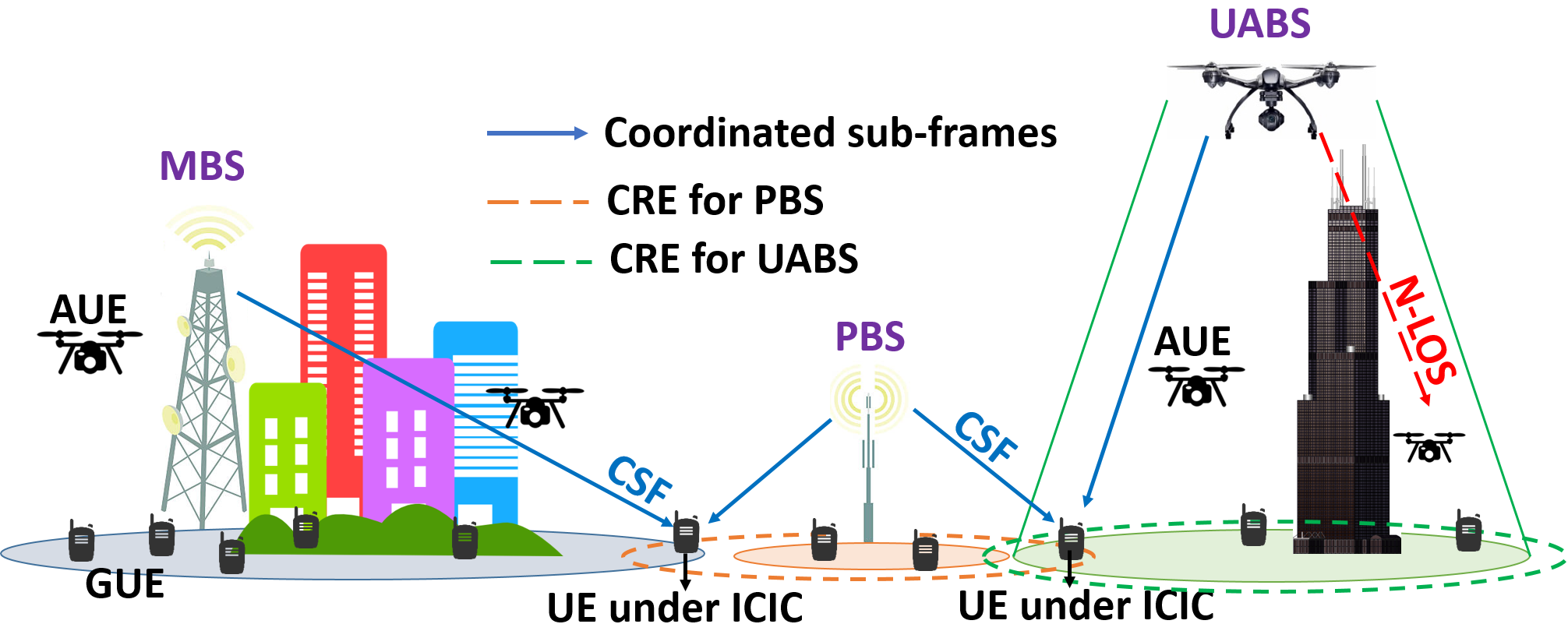}
\caption{The terrestrial nodes (MBS, PBS, and GUE) and aerial nodes (UABS and AUE) constitute the AG-HetNet.}
\label{PscHetnet}
\vspace{-6mm}
\end{figure}

Despite the earlier works given in~\cite{azari2017coexistence,kumbhar2018exploiting}, to the best of our knowledge, there are no prior works that consider both AUEs and UABSs to simultaneously co-exist with terrestrial nodes such as the macro base-stations (MBSs), pico base-stations (PBSs), and ground user equipment (GUEs) in LTE-Advanced AG-HetNet. To this purpose, we simulate an AG-HetNet in public safety band class~14 as shown in Fig.~\ref{PscHetnet}; which leverages on 3GPP Rel-8 cell range expansion (CRE), 3GPP Rel-10/11 intercell interference coordination (ICIC), 3GPP Rel-12 three-dimensional (3D) beamforming (3DBF), and 3GPP Rel-15 enhanced support for UAVs. Subsequently, we maximize the two KPIs of the network while mitigating intercell interference and jointly optimizing ICIC and CRE network parameters. Our simulation results show that a three-tier hierarchical structuring of reduced power subframes can effectively help mitigate interference in AG-HetNets. 

The rest of this paper is organized as follows. In Section~\ref{systemModel}, we provide the AG-HetNet system model, 3D channel model, 3DBF, and definition of KPIs as a function of network parameters. In Section~\ref{PerfAndSim}, we configure UABSs deployment on a hexagonal grid and present ICIC network parameters. In Section~\ref{simRes}, through extensive computer simulations, we analyze and compare the two KPIs of the AG-HetNet for various ICIC techniques and configurations. Finally, the last section provides concluding remarks.

\section{System Model}
\label{systemModel}
We consider a three-tier AG-HetNet deployment, where all the MBS, PBS and UABS locations (in 3D) are captured in matrices ${\bf X}_{\rm mbs} \in \mathbb{R}^{N_{\rm mbs}\times 3}$, ${\bf X}_{\rm pbs} \in \mathbb{R}^{N_{\rm pbs}\times 3}$, and ${\bf X}_{\rm uabs}\in \mathbb{R}^{N_{\rm uabs}\times 3}$, respectively, with $N_{\rm mbs}$, $N_{\rm pbs}$ and $N_{\rm uabs}$ denoting the number of MBSs, PBSs, and UABSs within the simulation area (${\rm A_{sim}}$).
Similarly, the 3D distribution of GUEs and AUEs are respectively captured in matrices ${\bf X}_{\rm gue}$ and ${\bf X}_{\rm aue}$. Assuming a fixed antenna height, the location of wireless nodes MBS, PBS, GUE, and AUE are modeled using a 2D Poisson point process (PPP), with intensities $\rm \lambda_{\rm mbs}$, $\rm \lambda_{\rm pbs}$, $\rm \lambda_{\rm gue}$ and $\rm \lambda_{\rm aue}$, respectively. On the other hand, UABSs are deployed on a fixed hexagonal grid and at two different heights (see also Table~\ref{tab:SysParams}).

For an arbitrary $n$th UE, let $d_{on}$, $d_{pn}$, and $d_{un}$ be the nearest distance from macrocell of interest (MOI), picocell of interest (MOI), and UABS-cell of interest (UOI), respectively. Then assuming Nakagami-m fading channel, the reference symbol received power from MOI, POI, and UOI is given by
\begin{align}
\label{eq:refPwr}  
& {R}_{\rm mbs}(d_{on}) = \frac{P_{\rm mbs}A_E(\phi, \theta)H}{10^{\varphi(d_{on})/10}},   
{R}_{\rm pbs}(d_{pn}) = \frac{P_{\rm pbs}A_E(\phi, \theta)H}{10^{\varphi(d_{pn})/10}}, \nonumber\\ 
& {R}_{\rm uabs}(d_{un}) = \frac{P_{\rm uabs}A_E(\phi, \theta)H}{10^{\varphi(d_{un})/10}},  
\end{align}
where random variable $H$ accounts for Nakagami-m fading and is defined in (2) of \cite{azari2017coexistence}. Through shaping parameter $m$, received signal power can be approximated to achieve variable fading conditions. The value $m > 1$ approximates to Rician fading along line-of-sight (LOS)  and $m = 1$ approximates to Rayleigh fading along non-LOS (NLOS). The variable $A_E(\phi,\theta)$ is the transmitter antenna's 3DBF element defined in (19)--(21) of \cite{kammoun2014preliminary}. Using 3DBF, the power transmission from MBS ($P_{\rm mbs}$), PBS ($P_{\rm pbs}$), and UABS ($P_{\rm uabs}$) can be controlled for UEs in cell-edge/CRE region. This limits the power transmission into adjacent cells that causes intercell interference and subsequently improves signal-to-interference ratio (SIR) of desired signal. The variables $\varphi(d_{on})$, $\varphi(d_{pn})$, and $\varphi(d_{un})$ are path-loss respectively observed from MBS, PBS, and UABS in dB.

\subsection{Path Loss Model}
Based on the type of communication link, i.e., ground-to-ground (GTG), any-to-air (ATA), and air-to-ground (ATG) between a UE and base-station (BS) of interest, we consider distinct path-loss models for an accurate analysis of signal reliability.

We consider Okumura-Hata path loss (OHPL) to estimate the GTG communication link between GUE and terrestrial MBS and PBS. OHPL in an urban terrestrial environment is defined in (1)--(2) of~\cite{xiroOnline}. In an urban-macro with aerial scenario, we consider ATA communication link between an AUE and any nearest BS. The average path loss for ATA link is calculated over the probabilities of LOS/NLOS defined in Table B-1, and path loss in Table B-2 of~\cite{3GPP.TR.36.777}. The average path loss for ATG communication link between GUE and UABS is calculated over the probabilities of LOS/NLOS defined in (4) of \cite{azari2017coexistence}.

\begin{figure}[h]
\centering{\includegraphics[width=0.7\linewidth]{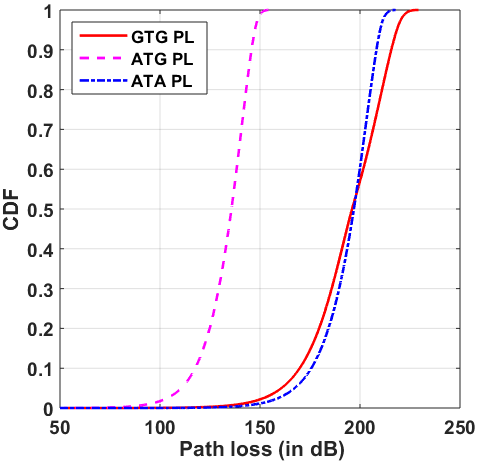}}
\caption{ The CDF of path loss observed for the communication link between UEs and base-stations.}
\label{fig:all3PL}
\end{figure}

Fig.~\ref{fig:all3PL} illustrates the empirical path loss cumulative distribution functions (CDFs), calculated for all distances between base stations (${\bf X}_{\rm mbs}$, ${\bf X}_{\rm pbs}$, and ${\bf X}_{\rm uabs}$) and UEs (${\bf X}_{\rm gue}$ and ${\bf X}_{\rm aue}$), using conditions defined in previous paragraph. Inspection of Fig.~\ref{fig:all3PL} reveals that the maximum allowable path loss is diverse for GTG, ATG, and ATA communication links. This variation is primarily due to the environmental factors and LOS/NLOS probability of communication link. Nevertheless, maximum allowable path-loss for the models used in GTG, ATA, and ATG link is approximately 255 dB, 216 dB, and 154 dB, respectively.

\begin{table*}[t]
\vspace{-3mm}
\caption{SIR and SE definitions.}
\label{tab:sirCap}
\centering
\begin{tabular}{p{5.25cm}|p{5.25cm}} 
\hline
      \textbf{Signal-to-interference ratio (SIR)} & \textbf{SE in USF/CSF radio frames}\\
      \hline
      $\Gamma^{\rm mbs} = \frac{R_{\rm mbs}(d_{on})}{R_{\rm pbs}(d_{pn}) + R_{\rm uabs}(d_{un}) + \mathbf{I}_{\rm agg}}$ & $C_{\rm usf}^{\rm mbs} = \frac{\beta_{\rm mbs} {\rm log_2}(1+\Gamma^{\rm mbs})}{N_{\rm usf}^{\rm mbs}}$\\
      $\Gamma^{\rm mbs}_{\rm csf} = \frac{\alpha R_{\rm mbs}(d_{on})}{\alpha_{\rm pbs}R_{\rm pbs}(d_{pn}) + R_{\rm uabs}(d_{un}) +  \mathbf{I}_{\rm agg}}$ & $C_{\rm csf}^{\rm mbs} = \frac{(1-\beta_{\rm mbs}){\rm log_2}(1+\Gamma^{\rm mbs}_{\rm csf})}{N_{\rm csf}^{\rm mbs}}$\\ \hline
      $\Gamma^{\rm pbs} = \frac{R_{\rm pbs}(d_{pn})}{R_{\rm mbs}(d_{on}) + R_{\rm uabs}(d_{un}) +  \mathbf{I}_{\rm agg}}$ & $C_{\rm usf}^{\rm pbs} = \frac{\beta_{\rm pbs} {\rm log_2}(1+\Gamma^{\rm pbs})}{N_{\rm usf}^{\rm pbs}}$\\
      $\Gamma^{\rm pbs}_{\rm csf} = \frac{\alpha_{\rm pbs} R_{\rm pbs}(d_{pn})}{\alpha R_{\rm mbs}(d_{on}) + R_{\rm uabs}(d_{un})+ \mathbf{I}_{\rm agg}}$ & $C_{\rm csf}^{\rm uabs} = \frac{(1 - \beta_{\rm pbs}) {\rm log_2}(1+\Gamma^{\rm uabs})}{N^{\rm pue}_{\rm usf}}$\\ \hline
      $\Gamma^{\rm uabs} = \frac{R_{\rm uabs}(d_{un})}{R_{\rm mbs}(d_{on}) + R_{\rm pbs}(d_{pn}) +  \mathbf{I}_{\rm agg}}$ & $C_{\rm usf}^{\rm mbs} = \frac{(\beta_{\rm mbs}+\beta_{\rm pbs}) {\rm log_2}(1+\Gamma^{\rm uabs})}{N_{\rm usf}^{\rm uue}}$\\
      $\Gamma^{\rm uabs}_{\rm csf} = \frac{R_{\rm uabs}(d_{un})}{\alpha R_{\rm mbs}(d_{on})+ \alpha_{\rm pbs}R_{\rm pbs}(d_{pn})+  \mathbf{I}_{\rm agg}}$ & $C_{\rm csf}^{\rm uabs} = \frac{(2-(\beta_{\rm mbs} + \beta_{\rm pbs})){\rm log_2}(1+\Gamma^{\rm uabs}_{\rm csf})}{N^{\rm uue}_{\rm csf}}$\\ \hline
    \end{tabular}
\vspace{-3mm}
\end{table*}

\subsection{Spectral Efficiency with 3GPP Rel.10/11 ICIC}
\label{icicidetails}
We consider CRE at small cells such as PBS and UABS to extend the network coverage and increase capacity, by offloading traffic from congested cells; nevertheless, an adverse side effect of CRE includes increased interference at UEs in cell-edge/CRE region.

To address this intercell interference, both MBS and PBS are capable of using 3GPP Rel-10/11 ICIC techniques, wherein  MBS and PBS can transmit radio frames at reduced power levels. The radio subframes with reduced power are termed as coordinated subframes (CSF) and full power as uncoordinated subframes (USF). The power reduction factor is given by $\alpha_{\rm mbs}$ and $\alpha_{\rm pbs}$ at MBS and PBS. In particular, $\alpha_{\rm mbs}=\alpha_{\rm pbs}=0$ corresponds to Rel-10 almost blank subframes eICIC, $\alpha_{\rm mbs}=\alpha_{\rm pbs}=1$ corresponds to no ICIC, and otherwise corresponds to Rel-11 reduced power FeICIC. We coordinate USF/CSF duty cycle using $\beta_{\rm mbs}$ and $(1-\beta_{\rm mbs})$ at MBS and $\beta_{\rm pbs}$ and $(1-\beta_{\rm pbs})$ at PBS.

\begin{figure}[t]
\vspace{-2mm}
\centering{\includegraphics[width=1\linewidth]{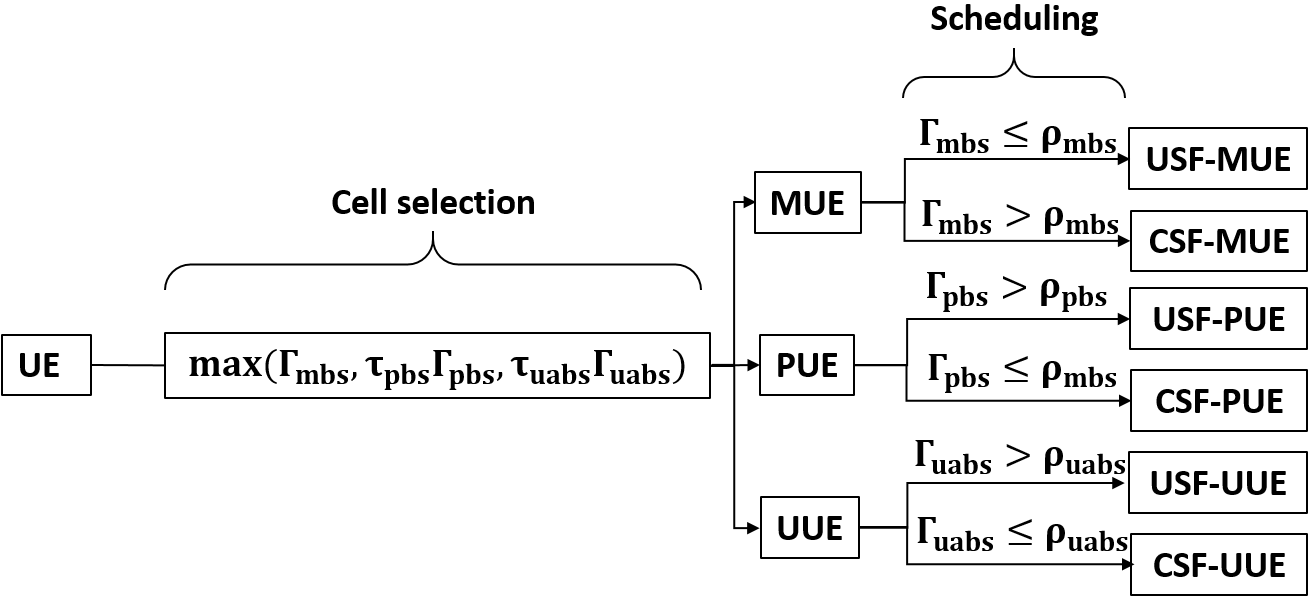}}
\caption{Cell selection and UE association in USF/CSF subframes of MBS, PBS, and UABS.}
\label{fig:CellSelection}
\vspace{-5mm}
\end{figure}

Let $\Gamma^{\rm mbs}$, $\Gamma^{\rm mbs}_{\rm csf}$, $\Gamma^{\rm pbs}$, $\Gamma^{\rm pbs}_{\rm csf}$, $\Gamma^{\rm uabs}$, and $\Gamma^{\rm uabs}_{\rm csf}$ denote SIR at USF and CSF subframes of MOI, POI, and UOI, respectively. Then, using positive biased CRE $\tau_{\rm pbs}$ at PBSs and $\tau_{\rm uabs}$ at UABSs, these small cells can expand their SIR coverage. Subsequently, during the process of cell selection, a UE always camps on the nearest BS that yields the best SIR. Then an individual MBS, or PBS, or UABS can schedule their UE in either USF/CSF radio subframes based on their respective scheduling threshold $\rho_{\rm mbs}$, $\rho_{\rm pbs}$, $\rho_{\rm uabs}$. This association of a UE with the nearest BS and scheduling in USF/CSF subframes for six different scenarios is summarized in Fig.~\ref{fig:CellSelection}.

By following an approach similar to that of~\cite{kumbhar2018exploiting}, we define the SIR and 5pSE experienced by an $n$th arbitrary UE for six different scenarios and are given in Table~\ref{tab:sirCap}. Therein, $\mathbf{I}_{\rm agg}$ is the aggregate interference at a UE from all the BSs, except from MOI, POI, and UOI, while $N_{\rm usf}^{\rm mue}$, $N_{\rm csf}^{\rm mue}$, $N_{\rm usf}^{\rm pue}$, $N_{\rm csf}^{\rm pue}$, $N^{\rm uue}_{\rm usf}$, and $N^{\rm uue}_{\rm csf}$ are the number of MBS-UE, PBS-UE, and UABS-UE scheduled in USF/CSF.

\section{Key Performance Indicators}
\label{PerfAndSim}
In this article, 5pSE corresponds to the worst fifth percentile UE capacity amongst all of the scheduled UEs. On the other hand, we define the coverage probability of the network as the percentage of an area having broadband rates and capacity larger than a threshold of~$T_{C_{\rm SE}}$.

In this study, we maximize the two KPIs of the network, while obtaining the best ICIC network configuration using a \textit{brute force algorithm}. However, the brute force algorithm is computationally infeasible to search for all possible optimal values in a large search space. Therefore, to reduce the system complexity and simulation runtime, we consider UABSs deployment on a fixed hexagonal grid and apply the same ICIC parameters across all MBSs, PBSs, and UABSs. With a feasible set of vectors, we determine the best state, $\bf{S^\prime}_{\mathbf{KPI}}$, out of all possible states $\bf{S}$ such that: 
\begin{equation}
\mathbf{S^\prime}_{\rm \mathbf{KPI}}  ={\rm \arg}~\underset{\mathbf{S}}{\rm \max} ~~ C_{\rm \mathbf{KPI}}(\mathbf{S}),  \label{Eq:optimizeState}
\end{equation}
where $\mathbf{KPI} \in \big(\rm 5pSE, COV\big)$. The objective function $C_{\rm 5pSE}(.)$ denotes 5pSE and $C_{\rm cov}(.)$ denotes coverage probability for a given state $\mathbf{S} = \Big[{\bf X}_{\rm uabs},{\bf S}_{\rm mbs}^{\rm ICIC},{\bf S}_{\rm pbs}^{\rm ICIC},{\bf S}_{\rm uabs}^{\rm ICIC}\Big]$. As defined previously, ${\bf X}_{\rm uabs}$ is the matrix representing the location of the $N_{\rm uabs}$ UABSs in three dimensions, ${\bf S}_{\rm mbs}^{\rm ICIC} = [\boldsymbol{\alpha_{\rm mbs}}, \boldsymbol{\beta_{\rm mbs}}, \boldsymbol{\rho_{\rm mbs}}]$ $\in \mathbb{R}^{N_{\rm mbs} \times 3}$ is a matrix that captures individual ICIC parameters for each MBS, ${\bf S}_{\rm pbs}^{\rm ICIC} = [\boldsymbol{\alpha_{\rm pbs}}, \boldsymbol{\beta_{\rm pbs}}, \boldsymbol{\rho_{\rm pbs}}, \boldsymbol{\tau_{\rm pbs}}]$ $\in \mathbb{R}^{N_{\rm pbs} \times 4}$ is a matrix that captures individual ICIC parameters for each PBS, and ${\bf S}_{\rm uabs}^{\rm ICIC} = [\boldsymbol{\tau_{\rm uabs}},\boldsymbol{\rho_{\rm uabs}}]$ $\in \mathbb{R}^{N_{\rm uabs} \times 2}$ is a matrix that occupies individual ICIC parameters for each UABS.

\begin{table}[t]
\caption{System and simulation parameters.}
\label{tab:SysParams}
\centering
\begin{tabular}{p{4.5cm} p{3cm}}
\hline
{\bf Parameter} & {\bf Value}  \\ \hline
Simulation area ($A_{\rm sim}$)  & $100 {\rm\ km^2}$\\ \hline
MBS, PBS, GUE, AUE intensities & 4, 12, 100, and 1.8 per km$^2$ \\ \hline
Number of UABS & 60 \\ \hline
MBS, PBS, and UABS transmit powers & 46, 30, and 26 dBm\\ \hline
Height of MBS, PBS, and UABS & 36 and 15m\\ \hline
Height of UABS & 36 and 50 m\\ \hline
Height of GUE and AUE & 1.5 and 22.5 m\\ \hline
PSC LTE Band~14 center frequency & 763 MHz for downlink\\ \hline
Power reduction factor $\alpha_{\rm mbs}$ and $\alpha_{\rm pbs}$ &  $0$ to $1$  \\ \hline
USF Duty cycle $\beta_{\rm mbs}$, $\beta_{\rm pbs}$ &  $0$ to $100$\%   \\ \hline
Scheduling threshold for MUEs ($\rho_{\rm mbs}$)            &  $20$ dB to $40$ dB \\ \hline
Scheduling threshold for PUEs ($\rho_{\rm pbs}$)            &  $-10$ dB to $10$ dB \\ \hline
Scheduling threshold for UUEs ($\rho_{\rm uabs}$)      &  $-5$ dB to $5$ dB \\ \hline
Range expansion bias for $\tau_{\rm uabs}$, $\tau_{\rm uabs}$ & $0$ dB to $12$ dB \\ \hline
\end{tabular}
\vspace{-3mm}
\end{table}

\section{Simulation Results}
\label{simRes}
\begin{figure}[t]
\centering
\begin{subfigure}[b]{0.235\textwidth}
\label{fig:HexCovProb}
\includegraphics[width=1.05\textwidth]{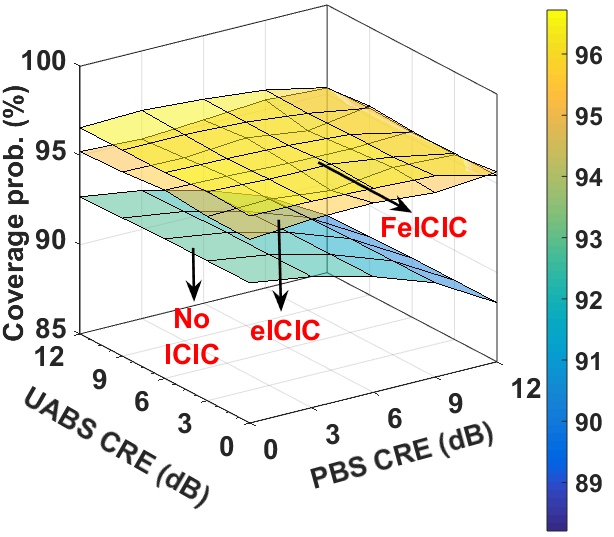}
\caption{Coverage prob. vs. CRE.}
\end{subfigure}
\begin{subfigure}[b]{0.235\textwidth}
\label{fig:Hex5pSE}
\includegraphics[width=1.09\textwidth]{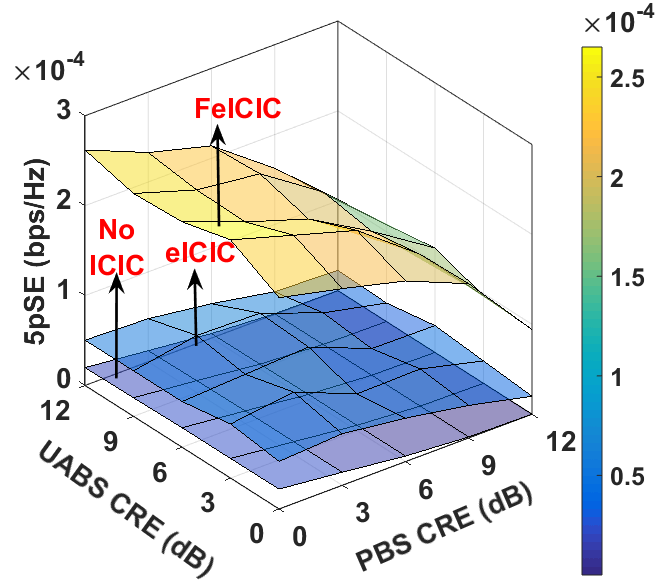}
\caption{Peak 5pSE vs. CRE.}
\end{subfigure}
\caption{The effects of combined CRE at PBS and UABS on the two KPIs of the network, with and without ICIC; when UABS are deployed at height of 36 m.}
\label{fig:HexKpiH36} 
\vspace{-3mm}
\end{figure}

\begin{figure}[t]
\centering
\begin{subfigure}[b]{0.235\textwidth}
\label{fig:HexCovProb}
\includegraphics[width=1.01\textwidth]{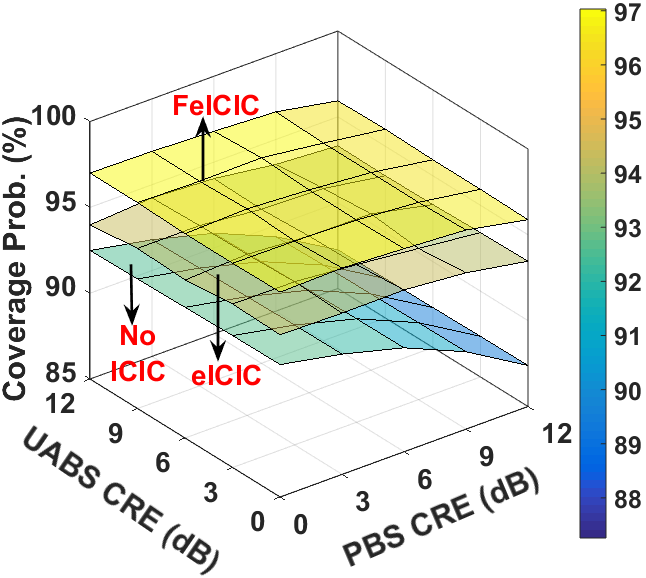}
\caption{Coverage prob. vs. CRE.}
\end{subfigure}
\begin{subfigure}[b]{0.235\textwidth}
\label{fig:Hex5pSE}
\includegraphics[width=1.09\textwidth]{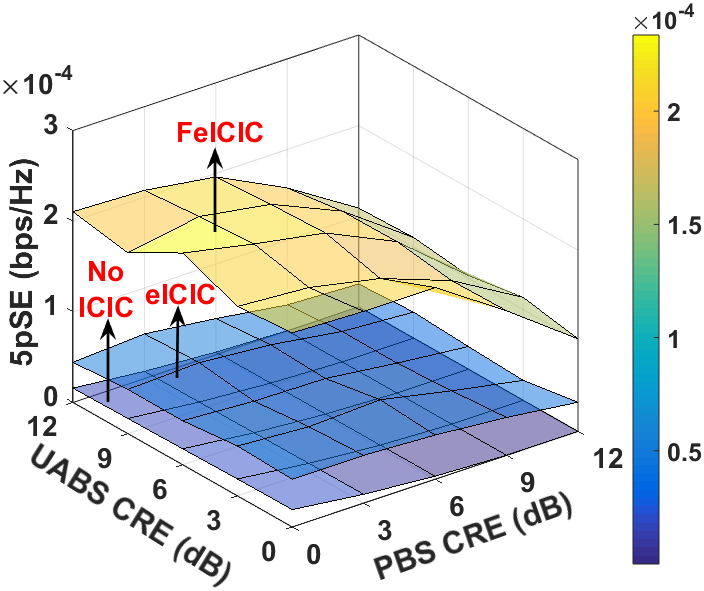}
\caption{Peak 5pSE vs. CRE.}
\end{subfigure}
\caption{The effects of combined CRE at PBS and UABS on the two KPIs of the network, with and without ICIC; when the UABS are deployed at height of 50 m.}
\label{fig:HexKpiH50} 
\vspace{-1mm}
\end{figure}

In this section, with the help of computer simulation and system parameters set to the values given in Table~\ref{tab:SysParams}, we compare the two KPIs of the network with and without ICIC techniques. The 3D surface plot in Fig.~\ref{fig:HexKpiH36} and Fig.~\ref{fig:HexKpiH50} illustrates the combined effect of CRE at PBSs and UABSs (along x- and y-axes) on the coverage probability and 5pSE (along the z-axis) of the wireless network. In an initial inspection of Fig.~\ref{fig:HexKpiH36} and Fig.~\ref{fig:HexKpiH50}, we can intuitively conclude that FeICIC performs better when compared to eICIC and without any ICIC techniques.

When UABS are deployed at the same height as MBS or height higher than MBS; in the absence of any ICIC mechanism, the optimal value of the CRE for coverage probability and 5pSE is observed at around 0 dB as seen in Fig.~\ref{fig:HexKpiH36} and Fig.~\ref{fig:HexKpiH50}. However, as the CRE increases, the interference also increases at scheduled UEs. As a result, the performance of the two KPIs starts to decline.

When the UABS are deployed at the same height as MBS, with eICIC, the two KPIs are seen to perform better when CRE at PBSs is between $3-6$ dB and at $3$ dB for UABSs. With FeICIC, the two KPIs are seen to perform better when CRE at PBSs is at $0$ dB and between $3-12$ dB for UABSs. Whereas, when we deploy UABS at a height higher than MBS, with eICIC, the two KPIs are seen to perform better when CRE is between $0-3$ dB for both PBSs and UABSs. With FeICIC, the two KPIs are seen to perform better when CRE at PBSs is between $3-6$ dB and between $6-12$ dB for UABSs.

\begin{figure}[t]
\centering{\includegraphics[width=0.75\linewidth]{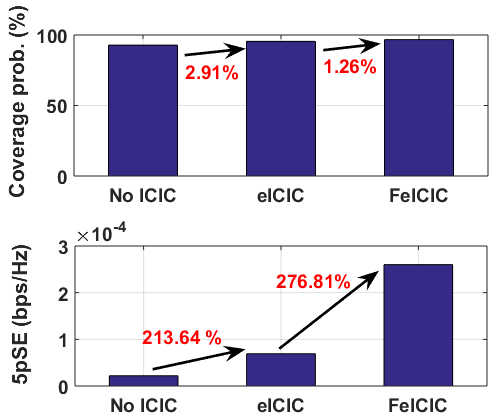}}
\caption{Performance comparison of the two KPIs; when UABS are deployed at height of 36 m.}
\label{fig:HexPerCompareH36}
\vspace{-5mm}
\end{figure}



The comparative analysis of Fig.~\ref{fig:HexKpiH36} and Fig.~\ref{fig:HexKpiH50} reveals that the improvement in coverage probability is less significant but the 5pSE improvement is significant. When UABS are deployed at the same height as MBS, for coverage probability eICIC sees an improvement of $2.91\%$ in the absence of any ICIC, and FeICIC sees an improvement of $1.26\%$ over eICIC. For 5pSE, eICIC sees an improvement of $213.64\%$ in the absence of any ICIC, and FeICIC sees an improvement of $276.81\%$ over eICIC. Whereas, when the UABS are deployed at a height higher than MBS, for coverage probability eICIC sees an improvement of $1.95\%$ in the absence of any ICIC, and FeICIC sees an improvement of $2.62\%$ over eICIC. For 5pSE, eICIC sees an improvement of $186.1\%$ in the absence of any ICIC, and FeICIC sees an improvement of $324.65\%$ over eICIC.

Finally, we also observe, as the deployment height of UABS increases, the LOS of UABS also increases.  As a result, interference at scheduled UEs increases, and there is a sparse decrease in the KPI values of the wireless network.

\begin{figure}[t]
\centering{\includegraphics[width=0.75\linewidth]{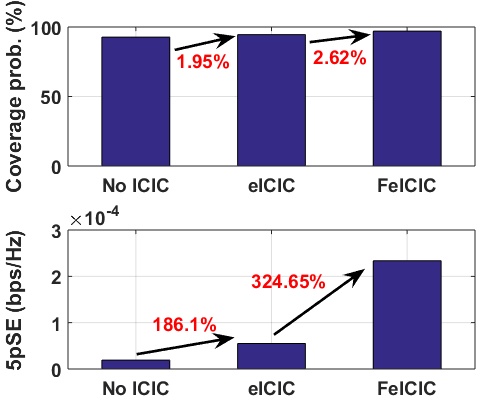}}
\caption{Performance comparison of the two KPIs; when the UABS are deployed at height of 50 m.}
\label{fig:HexPerCompareH50}
\vspace{-5mm}
\end{figure}

\section{Conclusion}
This paper gives system-level insights into the LTE-Advanced AG-HetNet. Through simulations, we maximized the coverage probability and 5pSE of the network, while addressing the intercell interference and optimizing the ICIC network parameters using a brute force technique. Our analysis shows that the HetNet with reduced power subframes (FeICIC) yields better coverage probability and 5pSE that than with almost blank subframes (eICIC).



\bibliographystyle{IEEEtran}
\bibliography{IEEEabrv,Citations}

\end{document}